\documentstyle[prd,preprint,aps,epsfig,floats]{revtex}
\draft
\newcommand{\beq}{\begin{equation}}
\newcommand{\eeq}{\end{equation}}
\newcommand{\bea}{\begin{eqnarray}}
\newcommand{\eea}{\end{eqnarray}}
\begin{document}

\title{Deeply Virtual Compton Scattering at HERA - A Probe of Asymptotia}
 
\author{L.L.Frankfurt$^a$, A.Freund$^b$, M. Strikman$^b$}

\address{
$^a$Physics Department, Tel Aviv University, Tel Aviv, Israel\\ 
$^b$Department of Physics, Penn State University\\
University Park, PA  16802, U.S.A.}

\maketitle

\begin{abstract}
We demonstrate that the measurement of an azimuthal angle asymmetry in 
deeply virtual Compton scattering (DVCS) at HERA energies, is 
experimentally feasible and allows one to determine for the 
first time the ratio, $\eta$, of the real to imaginary part of the DIS 
amplitude. We further show that such measurements would discriminate between 
different scenarios for the energy dependence of $F_2(x,Q^2)$ at energies 
beyond those reachable at HERA.

\end{abstract}

\section{Introduction}
\label{intro1}

It is generally agreed that the $x$-range currently available at HERA 
is not sufficient to test the current ideas about the onset of 
asymptotia via measurements of the parton densities.
Therefore, the aim of this paper is to draw attention to the fact that 
the derivative of 
parton distributions with respect to $\ln x$, which can be measured at HERA,
is rather sensitive to the asymptotic behaviour of parton densities at
$x\rightarrow 0$ which can be probed at the LHC only.
Actually, the experience in studies of soft processes tells us that
the  real part of the zero angle scattering amplitude, provides us, 
through the dispersion representation with respect to the invariant energy of 
the collision, with information about the energy dependence of the cross 
section well beyond the energy where real part of amplitude is measured. 
The reason for this is that 
$\eta$, the ratio of the real to imaginary part of the amplitude essentially 
measures the $\ln s$ derivative of the cross section \cite{1a}:
\beq
\eta = \frac{\pi}{2}\frac{d\ln(F_2(x,Q^2))}{d\ln(1/x)}.
\label{reimx}
\eeq
One can also use 
analyticity relations to derive a more accurate
formula \cite{bronzan}, leading to 
\beq
\eta = \frac{s^{\alpha}}{ImA(s,t)}tan\left[\frac{\pi}{2}\left
(\alpha -1 +\frac{d}{d\ln s}\right )\right]\frac{ImA(s,t)}{s^{\alpha}}.
\label{reimbr}
\eeq
for $F_2(x,Q^2) \propto x^{-\alpha}$.

We propose a new methodology for investigating the energy dependence of
hard processes through the real part of the amplitudes of high energy processes
and also through the shapes of nondiagonal parton distributions. DVCS offers 
us a direct way to study of nondiagonal parton distributions.
The idea is that at sufficiently small $x$ the difference between 
diagonal and off-diagonal effects influences the $x$
dependence of parton distributions only weakly. This has been known for a long
time from calculations of Regge pole behaviour in quantum field theory.
We also check that this statement is valid within the DGLAP
approximation. Thus DVCS can be used to investigate asymptotia of
parton distribution through the real part of the amplitude for DVCS.

Note that from a mathematical point of view, the actual extraction of 
nondiagonal parton distributions with the
help of a factorization theorem from the data is not possible in DVCS due to 
the fact that the parton 
distributions depend on $y_1$ and $y_2=y_1-x$ which are dependent 
variables rather than independent as one would need and thus the inverse 
Mellin transform
of the factorization formula cannot be found\footnote{This is not true for 
diffractive di-muon production since there, we have two independent variables 
$x$ and $\xi_1$, the longitudinal momentum fraction of the produced di-muon.}.
However, in practice, one will be able to neglect the dependence on $y_2$ at
sufficiently small $x$ and by encoding the difference in the evolution of 
nondiagonal to diagonal distribution in a $Q$-dependent function, one can 
indeed extract the nondiagonal parton distribution at small $x$ with an 
uncertainty associated with the $Q$ dependent function.

The major new result of our analysis is that the current successful fits 
to the $F_{2N}(x,Q^2)$ HERA data lead to qualitatively different predictions  
for the asymmetry, reflecting different
underlying assumptions of the fits about the behavior of parton densities
at $x$ below the HERA range. A recent analysis in Ref.\  \cite{1} has shown 
that DVCS studies at HERA are feasible and we made predictions for the 
expected DVCS counting rate compared to DIS as well as the asymmetry $A$ in 
the combined DVCS and Bethe-Heitler cross section for recent H1 data.

The paper is structured as follows. In Sec.\ \ref{basics} we review the 
necessary formulas of Ref.\ \cite{1} for our analysis. In this context, 
the formula pertaining to the ratio of real to imaginary part of a scattering 
amplitude at small $x$ is of particular importance. We then
present the different fits to $F_2(x,Q^2)$ in Sec.\ \ref{fits} and present
the different results for the asymmetry $A$ with respect to $t$ and $y$, at
fixed $y$ and $t$ respectively. Sec.\ \ref{concla} 
contains our conclusions and outlook.

\section{Relations between DVCS and DIS}
\label{basics}

In order to compute the asymmetry $A$, we need the ratio of the 
imaginary part of the DIS amplitude to the imaginary part of the DVCS 
amplitude and the relative DVCS counting rate $R_{\gamma}$, expected at HERA 
in the interesting kinematic regime of $10^{-4}<x<10^{-2}$ and moderate $Q^2$, 
i.e.\ , $3.5~\mbox{GeV}^2<Q^2<45~\mbox{GeV}^2$. The relative counting rate 
$R_{\gamma}$ is given by \cite{1} 
\beq
R_{\gamma} \simeq \frac{\pi \alpha}{4 R^2 Q^2 B}F_2(x,Q^2)(1 + \eta^2).
\label{fNa}
\eeq
where $R$ is the ratio of the imaginary parts of the DIS to DVCS amplitude
as given in \cite{1}\footnote{We will use the results for $R$ from \cite{1} in 
our present analysis.}, $B$ is the slope of the $t$ dependence 
(for more details see Ref.\ \cite{1}.) and $\eta$ is the ratio of real to 
imaginary part of the DIS amplitude, i.e.\ , $F_2(x,Q^2)$, given by 
Eq.\ (\ref{reimx}).

\begin{figure}
\centering
\mbox{\epsfig{file=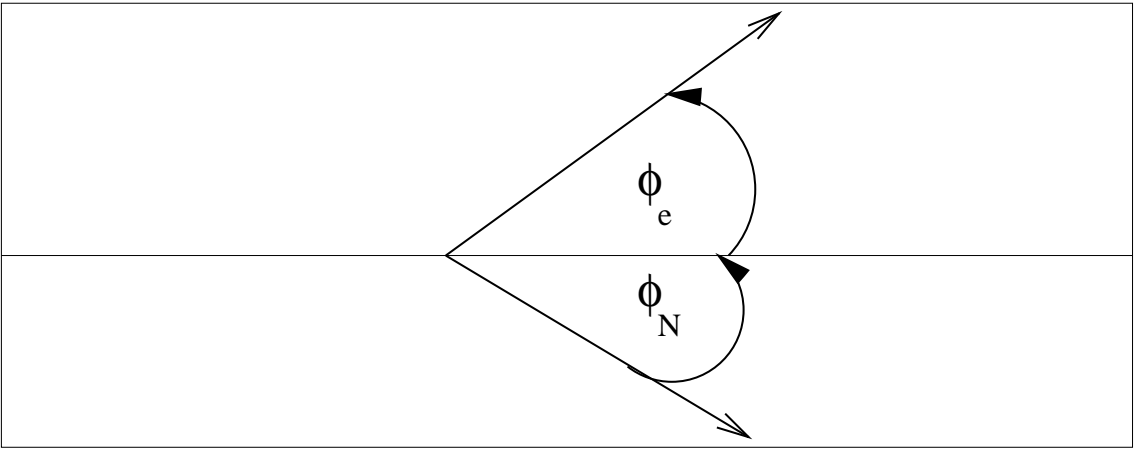,height=6cm}}
\vspace*{5mm}
\caption{The azimuthal final proton and electron angle in the transverse 
scattering plane.}
\label{angle}
\end{figure}

We also need the differential cross section for DVCS which can be simply 
expressed through the DIS differential cross section by multiplying the
DIS differential cross section by $R_{\gamma}$ (see Ref.\ \cite{1} for more
details.)
We then find using Eq.\ (\ref{fNa}) for $R_{\gamma}$
\beq
\frac{d\sigma_{DVCS}}{dxdyd|t|d\phi_r}=\frac{\pi\alpha^3s}{4R^2Q^6}(1+(1-y)^2)
e^{-B|t|}F^2_2(x,Q^2)(1+\eta^2)
\label{dvcsc}
\eeq
with $\sigma_{DVCS} = \frac{d\sigma_{DVCS}}{dt}|_{t=0}
\times \frac{1}{B}$. $B$ is the slope of the t dependence which we took to be
an exponential for simplicity. In writing Eq.\ (\ref{dvcsc}) we neglected 
$F_L(x,Q^2)$ - the experimentally observed conservation of s channel 
helicities in forward scattering high energy processes
justifies this approximation - so that $F_2\simeq 2xF_1$.
$y=1-E'/E$ where $E'$ ist the energy of the 
electron in the final state and $\phi_r=\phi_N + \phi_e$, 
where $\phi_N$ is the azimuthal angle between the plane defined by 
$\gamma^*$ and the final state proton and the $x-z$ plane and $\phi_e$
is the azimuthal angle between the plane defined by the initial and final state
electron and the $x-z$ plane (see Fig.\ \ref{angle}). Thus $\phi_r$ is nothing but
the angle between the $\gamma^*-p'$ and the electron's scattering planes.
 
In case of the Bethe-Heitler process, we find the differential cross section at
small $t$ to be
\bea
\frac{d\sigma_{BH}}{dxdyd|t|d\phi_r} &=& \frac{\alpha^3 s y^2(1+(1-y)^2)}{\pi Q^4|t| (1-y)}\times \left [ \frac{G_E^2(t) + \tau G_M^2(t)}{1+\tau} \right ]
\label{difcros}
\eea
with $\tau = |t|/4m_N^2$, $s$ being the invariant energy and $y$ the fraction of 
the scattered electron/positron energy.
$G_E(t)$ and 
$G_M(t)$ are the electric and nucleon form factors and we describe them using 
the dipole fit
\beq
G_E(t)\simeq G_D(t)=(1+\frac{|t|}{0.71})^{-2}~~\mbox{and}~~G_M(t)=\mu_p G_D(t),
\label{nform}
\eeq
where  $\mu_p=2.7$ is the proton magnetic moment.  
We make the standard assumption that the spin flip term is small in the 
strong amplitude for small $t$.

In order to write down the complete total cross section of exclusive photon 
production we need the interference term between DVCS and Bethe-Heitler. Note
that in the case of the interference term one does not have a spinflip in the 
Bethe-Heitler amplitude, i.e.\ , one only has $F_1(t)$, as compared to 
Eq.\ \ref{difcros} containing a spinflip part, i.e.\ , $F_2(t)$. The 
appropriate combination of $G_E(t)$ and $G_M(t)$ which yields $F_1(t)$ is  
\beq
\left[\frac{G_E(t) + \frac{|t|}{4m_N^2}G_M(t)}{1+\frac{|t|}{4m_N^2}}\right ].
\eeq

We then find for the interference term of the differential cross section, 
where we already use Eq.\ \ref{difcros}, 
\bea
\frac{d\sigma_{DVCS+BH}^{int}}{dxdyd|t|d\phi_r} &=& \pm \frac{\eta 
\alpha^3 s y(1+(1-y)^2) cos(\phi_r) e^{-B|t|/2} F_2(x,Q^2)}{2 Q^5 \sqrt(|t|)
\sqrt(1-y) R}\nonumber\\
& & \times \left [ \frac{G_E(t) + \tau G_M(t)}{1+\tau} \right ]
\label{inter}
\eea 
with the + sign corresponding to electron scattering off a proton and the - 
sign corresponding to the positron. The total cross section is then just the 
sum of Eq.\ \ref{dvcsc},\ref{difcros} and\ \ref{inter}.

We define the asymmetry $A$ as \cite{1}
\beq
A =\frac{\int_{-\pi/2}^{\pi/2}d\phi_r d\sigma_{DVCS+BH} - \int^{3\pi/2}_{\pi/2}
d\phi_r d\sigma_{DVCS+BH}}{\int_{0}^{2\pi}d\phi_r d\sigma_{DVCS+BH}},
\label{asym}
\eeq 
where $d\sigma_{DVCS+BH}$ is given by the sum of 
Eq.\ (\ref{dvcsc}),(\ref{difcros}),(\ref{inter}).
As explained in \cite{1} this azimuthal angle asymmetry is due to the fact
that the interference term in the combined DVCS and Bethe-Heitler cross section
is $\propto \frac{p_{t}}{\epsilon}$. Here $p_t$ is the component of the final 
proton
momentum transverse to the momentum of the initial electron and proton with
$\epsilon$ being polarization of the produced photon.
Integrating over the upper hemisphere, from $-\pi/2$ to $\pi/2$, one
obtains a $+$ sign from the intereference term and a $-$ sign from integrating
over the lower hemisphere of the detector, from $\pi/2$ to $3\pi/2$.

The real part of the DVCS amplitude is isolated through this asymmetry.
Therefore, we investigate the influence of different $F_2$ fits on the 
asymmetry through the relative counting rate which is directly sensitive to 
the ratio of real to imaginary parts of $F_2$ as shown in Eq.\ (\ref{fNa}).

\section{The different fits to $F_2(x,Q^2)$}
\label{fits}

In the calculation of the asymmetry $A$ we use the recent H1 data from Ref.\ 
\cite{10} as previously used in Ref.\ \cite{1}, a logarithmic fit by 
Buchm\"uller and Haidt (BH) \cite{11}, the ALLM97 fit \cite{12} and a leading 
order BFKL-fit \cite{13} for illustrative purposes.

In the H1 data, $F_2$ behaves for small $x$ as $x^{-\lambda}$ 
and hence $\eta$
is just $\frac{\pi}{2}\lambda$ where $\eta^2 = 0.09 - 0.27$ in the $Q^2$ range
given in the previous section. Note that $\eta$ has no $x$ dependence, 
for small enough $x$, and thus depends only on $Q^2$. This is not true for 
all of the other fits.

$F_2$ in the BH fit \footnote{In a more recent fit Haidt 
\cite{haidt} also used a double logarithmic fit with 
$log\left(\frac{Q^2}{Q^2_0}\right)\rightarrow log\left(1+\frac{Q^2}{Q^2_0}
\right)$ being the essential difference, save some minor adjustments for some 
constants, in order to be able to describe more recent low $Q^2$ data from 
HERA \cite{hera}. In the $Q^2$ range considered in this analysis the 
difference is negligible.} takes on the following form
\beq
F_2(x,Q^2) = 0.078 + 0.364\log(\frac{Q^2}{0.5~\mbox{GeV}^2})\log(\frac{0.074}
{x}),
\label{buf2}
\eeq
and hence we find for $\eta$
\beq 
\eta =\frac{\pi}{2}0.364\frac{\log(\frac{Q^2}{0.5~\mbox{GeV}^2})}{F_2(x,Q^2)}.
\label{bueta}
\eeq
Note that this $\eta$ has not only the usual $Q^2$ dependence but depends 
rather strongly on $x$ also.

In the ALLM97 fit $F_2$ at small $x$ takes on the following form
\beq
F_2(x,Q^2)=\frac{Q^2}{Q^2+m_0^2}(F_s^p(x,Q^2)+F_2^R(x,Q^2)),
\label{allmf2}
\eeq
where $\eta$ is then given by
\beq
\eta = -\frac{\pi}{2}\frac{a_Pc_Px_P^{a_P}+a_Rc_Rx_R^{a_R}}
{c_Px_P^{a_P}+c_Rx_R^{a_R}}.
\label{etaallm}
\eeq  
The different variables and constants used in the fit can be found in 
\cite{12}.

In the case of the leading order BFKL approximation where 
$F_2\simeq x^{-\frac{4N_c\ln(2)\alpha_s}{\pi}}$, we find $\eta$ to be
\beq
\eta = \frac{\pi}{2}\frac{4N_c\ln(2)\alpha_s}{\pi}.
\label{etabfkl}
\eeq

\section{Results for the asymmetry $A$}
\label{results}

\begin{figure}
\vskip-1in
\centering
\epsfig{file=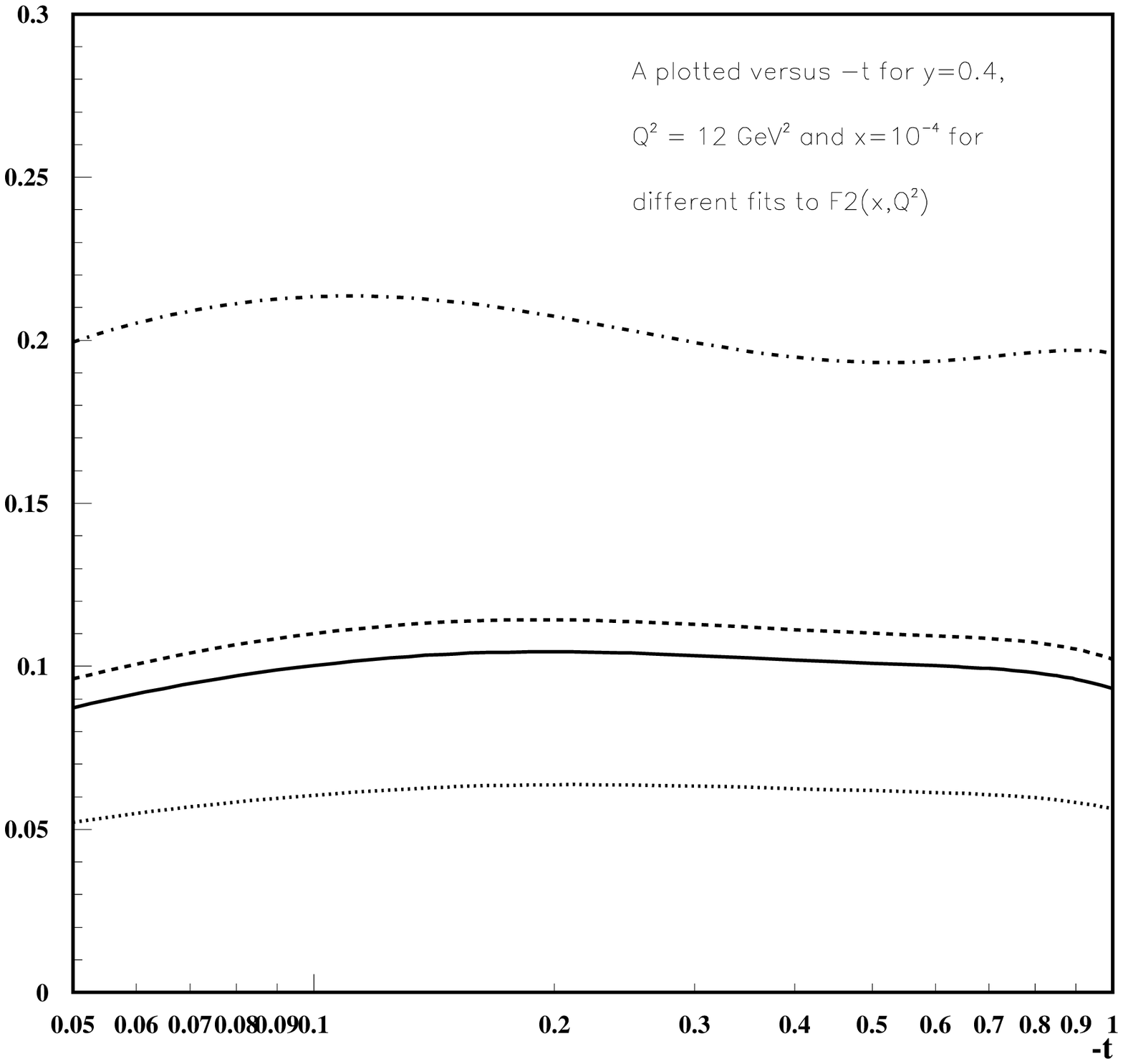,height=12cm}
\vskip-1in
\epsfig{file=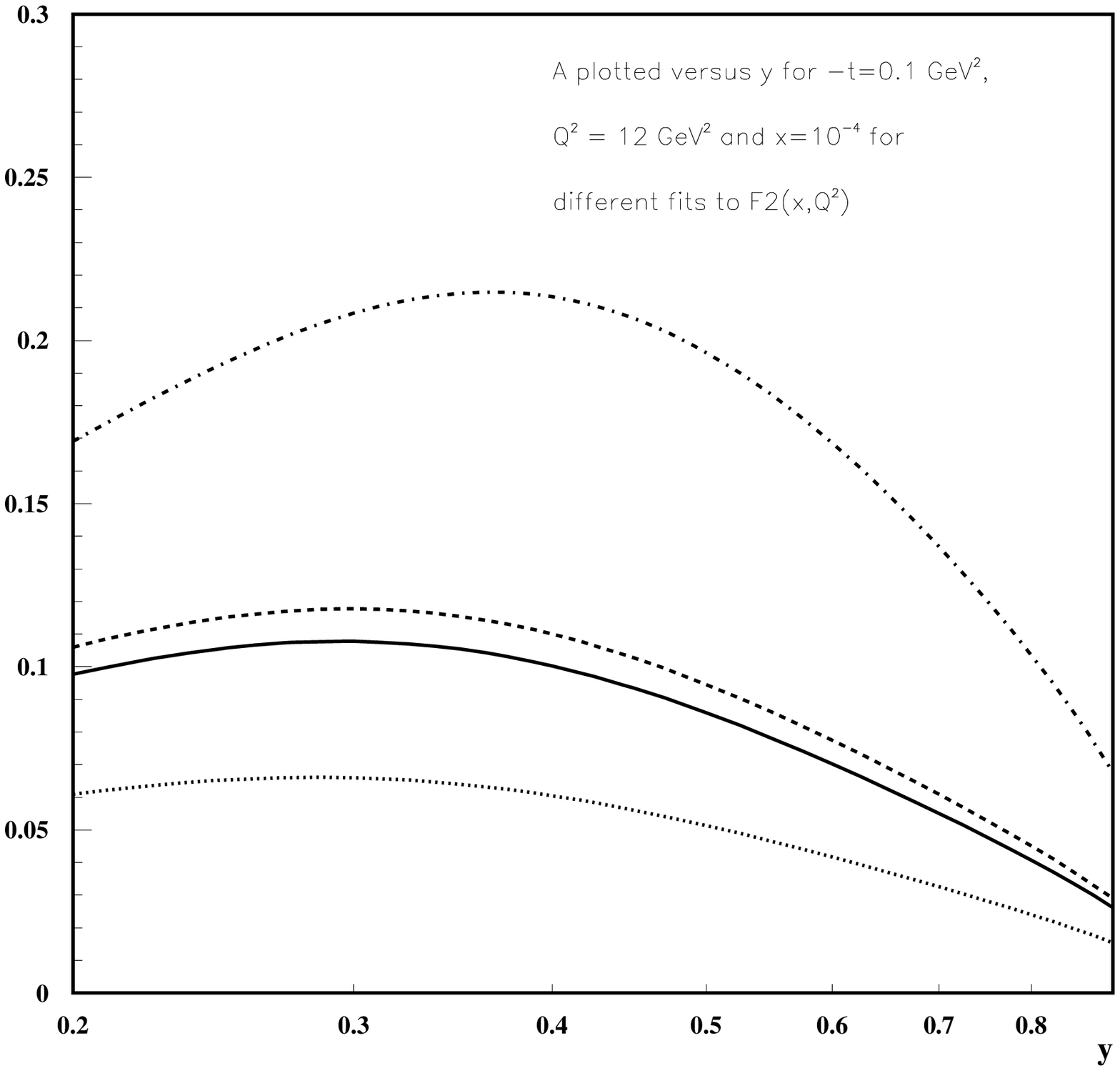,height=12cm}
\caption{H1 fit (solid curve), the BH 
fit (dotted curve), ALLM97 fit (short-dash curve) and BFKL fit (dash-dot 
curve)
for $x=10^{-4}$. a) Asymmetry $A$ versus $t$ for fixed $y=0.4$.
b) Asymmetry $A$ versus $y$ for fixed $-t=0.1~\mbox{GeV}^{2}$.}
\label{fig1}
\end{figure}

\begin{figure}
\vskip-1in
\centering
\epsfig{file=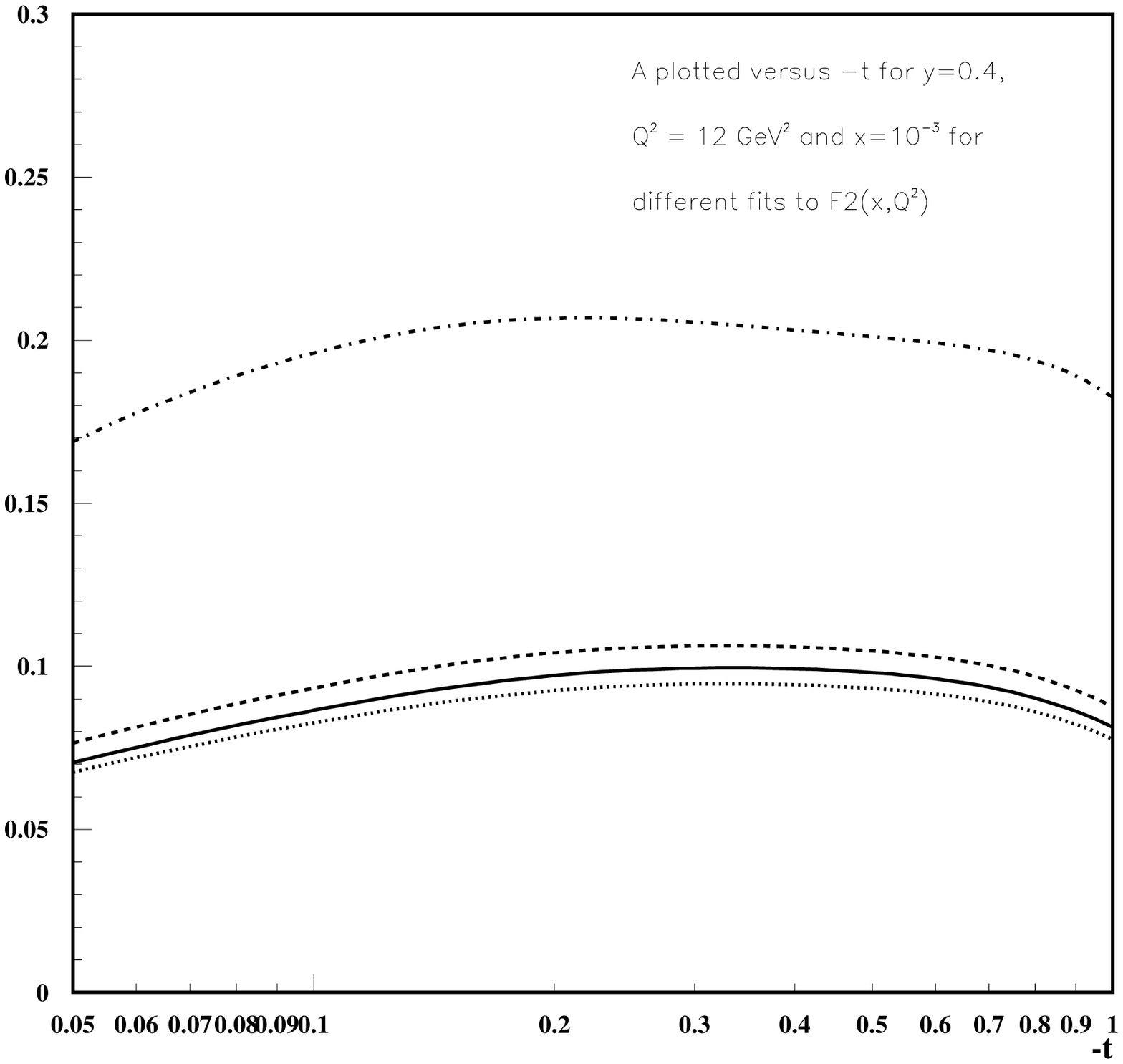,height=12cm}
\vskip-1in
\epsfig{file=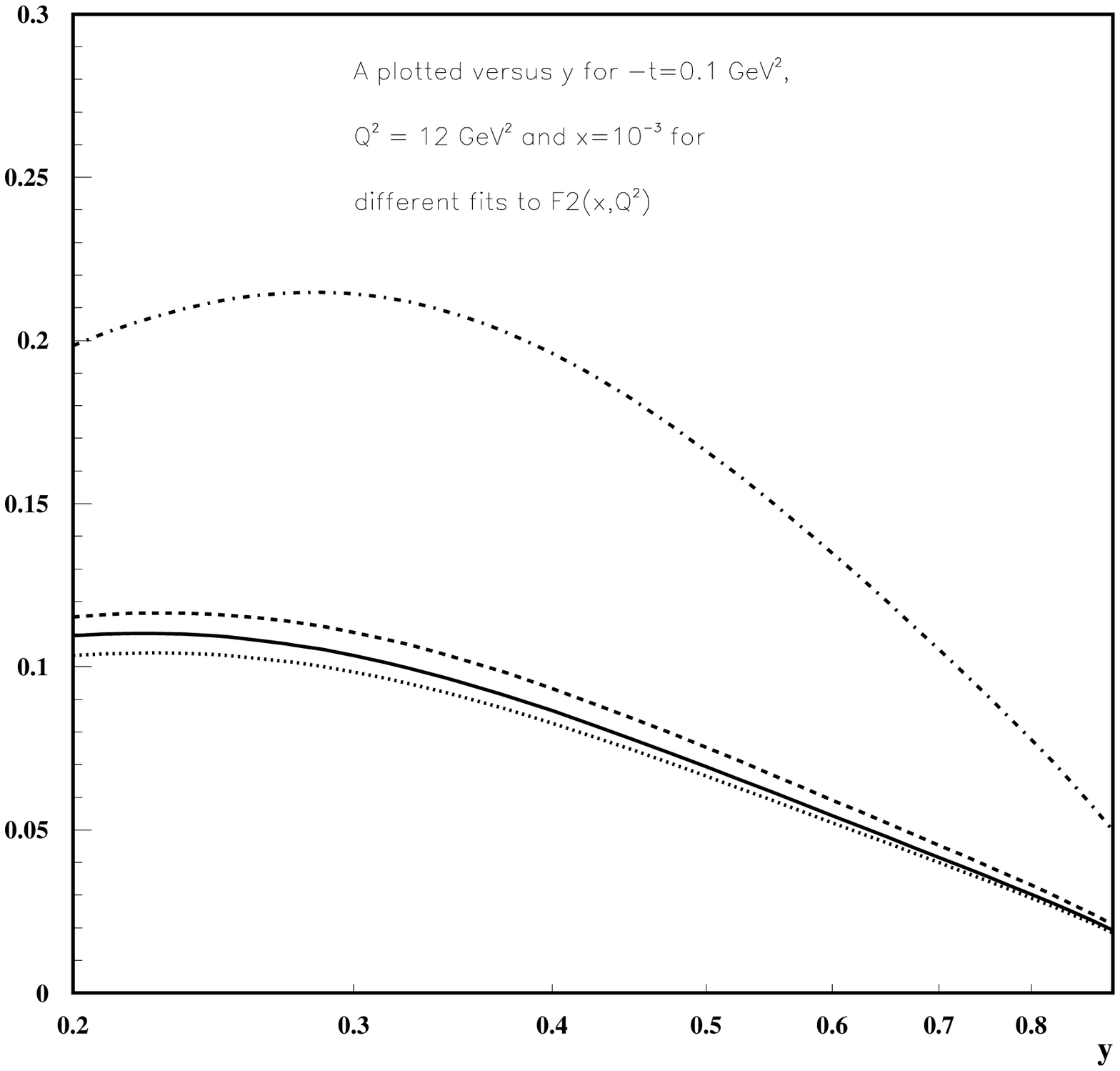,height=12cm}
\caption{H1 fit (solid curve), the BH 
fit (dotted curve), ALLM97 fit (short-dash curve) and BFKL fit (dash-dot 
curve)
for $x=10^{-3}$. a) Asymmetry $A$ versus $t$ for fixed $y=0.4$.
b) Asymmetry $A$ versus $y$ for fixed $-t=0.1~\mbox{GeV}^{2}$}
\label{fig2}
\end{figure}

\begin{figure}
\vskip-1in
\centering
\epsfig{file=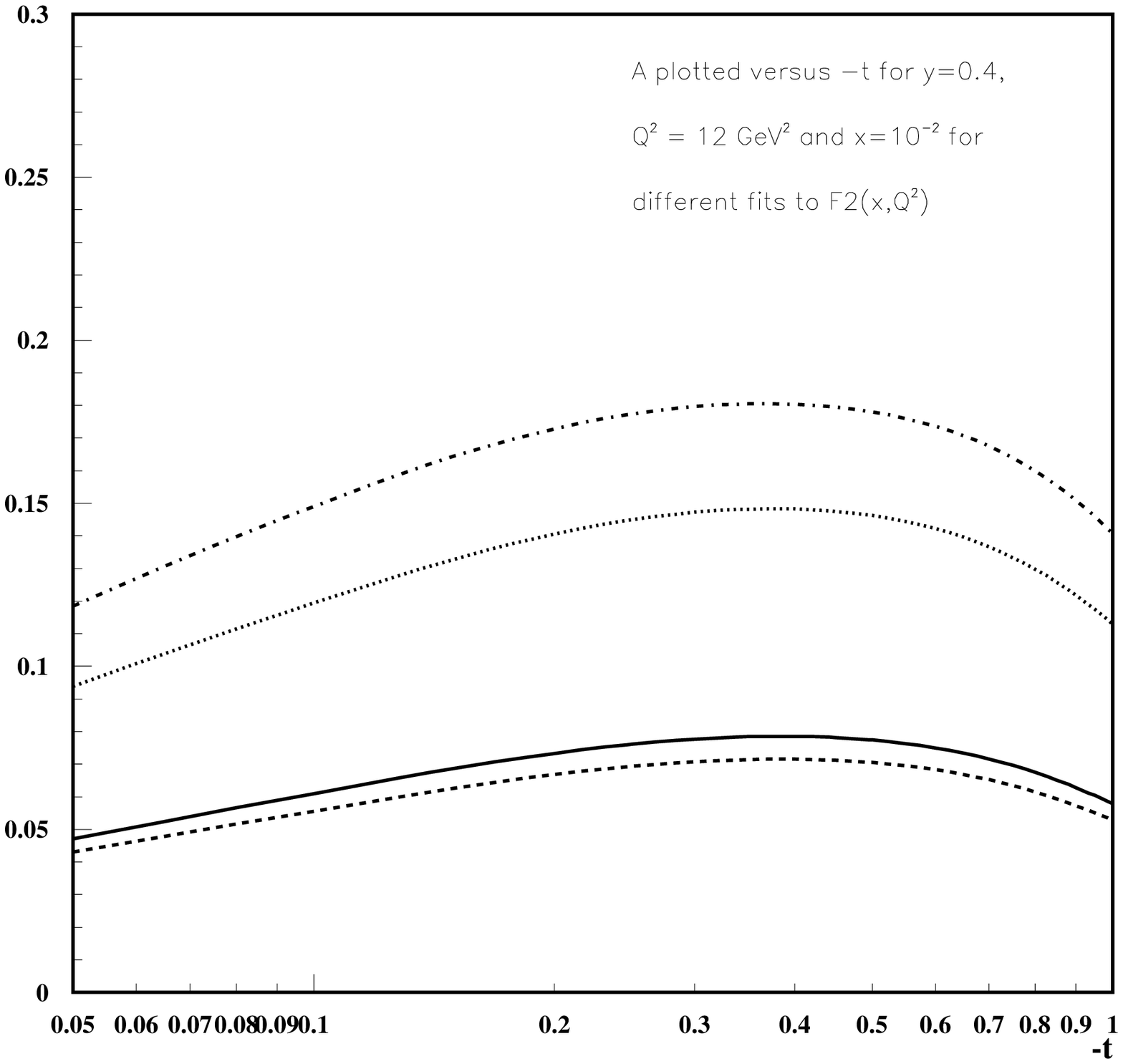,height=12cm}
\vskip-1in
\epsfig{file=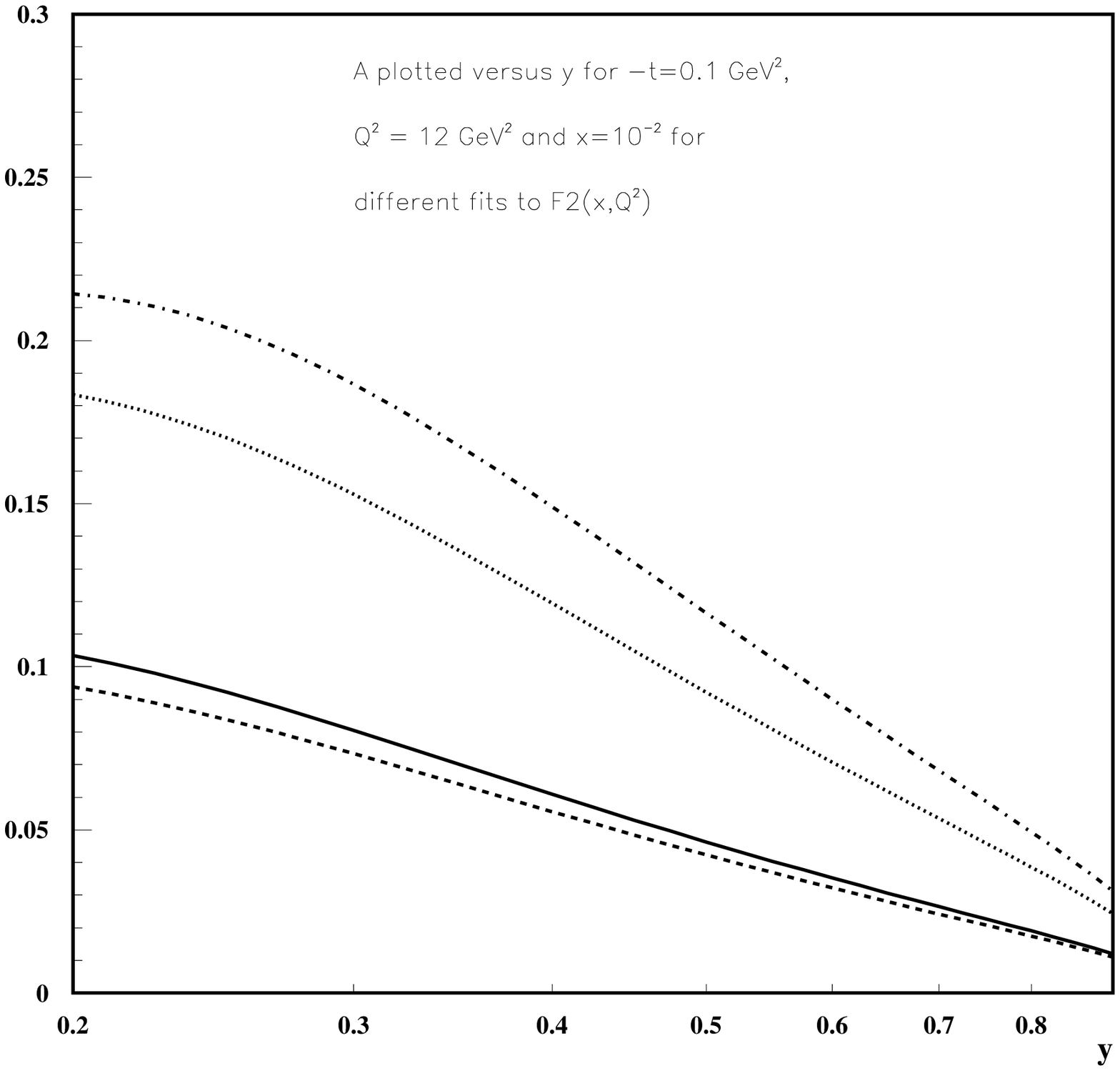,height=12cm}
\caption{H1 fit (solid curve), the BH 
fit (dotted), ALLM97 fit (short-dash curve) and BFKL fit (dash-dot 
curve)
for $x=10^{-2}$. a) Asymmetry $A$ versus $t$ for fixed $y=0.4$.
b) Asymmetry $A$ versus $y$ for fixed $-t=0.1~\mbox{GeV}^{2}$ }
\label{fig3}
\end{figure}

In Fig.\ \ref{fig1} - \ref{fig3}, we plot the asymmetry $A$ as a function of 
$t$ and $y$ for fixed 
$Q^2=12~\mbox{GeV}^2$, fixed $y=0.4$ and $-t=0.1~\mbox{GeV}^{2}$ and 
$x=10^{-4},~10^{-3},~10^{-2}$. The slope $B$ of the $t$-dependence for DVCS 
was taken to be 
$B=5~\mbox{GeV}^{-2}$ whereas for the Bethe-Heitler cross section we used the 
nucleon form factor as used in chapter 5. The counting rate $R_{\gamma}$ was 
appropriately adjusted for the different fits according to Eq.\ (\ref{fNa}).
The solid curves in Fig.\ \ref{fig1} - \ref{fig3} are our benchmarks\footnote
{Though actual H1 data is used, we are still dealing with a leading order
approximation and a particular model for the nondiagonal parton distributions
at the normalization point was used in computing $R_{\gamma}$(see \cite{1} 
for more details on the type of model ansatz and approximations used.).}.

Comparing the BH fit (medium-dashed curves), against our benchmarks 
we find a strong $x$ dependence of the asymmetry in the 
BH fit as well as different shapes and absolute values.

As far as the ALLM97 fit is concerned (short-dashed curves), there is hardly a 
difference, as compared to the H1 fit in the asymmetry as a function 
of $t$ and $y$
in absolute value, shape and $x$ dependence, except for $x=10^{-2}$ but this 
is due to the approximations we made for $x_P$ and $x_R$ which are not that 
good anymore at $x=10^{-2}$. 

If one compares the LO BFKL fit (long-dashed curves) to the H1 fit one sees 
immediately that the BFKL fit is totally off in almost all aspects and was 
only included here as an illustrative example.

\section{Conclusions}
\label{concla}

In the above we have shown the sensitivity of the exclusive
DVCS asymmetry $A$ to different $F_2$ fits and made comments on the 
viability of each fit. Note that even a fit which reproduces $F_2$ data, 
as well as its slope, in a satisfactory manner can be shown to lead to 
differences in the asymmetry shape. The sensitivity of the asymmetry to $y$ 
and $t$  will allow us, once 
experimentally determined, to make a shape fit and hence make a shape fit to 
nondiagonal parton distributions for the first time.

\section*{Acknowledgments}

This work was supported under DOE grant number DE-FG02-93ER40771.

\newpage

\end{document}